\documentclass[11pt]{article}

\usepackage{amsmath,amssymb}
\usepackage{slashed}
\usepackage{xcolor} 
\usepackage{graphicx}
\usepackage{dcolumn} 
\usepackage{bm} 
\usepackage{epsfig}
\usepackage{epstopdf}
\usepackage{grffile}
\usepackage{color}
\usepackage{colordvi}
\usepackage{amsmath,amssymb}
\usepackage{rotating}
\usepackage{lscape}
\usepackage{cite}
\usepackage{float}
\usepackage{hyperref}
\usepackage{url}
\usepackage{graphicx}
\usepackage{color}
\usepackage{amssymb}
\usepackage{amsmath}
\usepackage{mathtools}
\usepackage{bm}	
\usepackage{flexisym}
\usepackage{amssymb}
\usepackage{physics}
\usepackage{mathrsfs}
\usepackage{simplewick}
\usepackage{tikz}
\usetikzlibrary{arrows,shapes}
\usetikzlibrary{trees}
\usetikzlibrary{matrix,arrows} 			
\usetikzlibrary{positioning}			
\usetikzlibrary{calc,through}			
\usetikzlibrary{decorations.pathreplacing}  
\usepackage{pgffor}						
\usetikzlibrary{decorations.pathmorphing}
\usetikzlibrary{decorations.markings}
\tikzset{
	vector/.style={decorate, decoration={snake}, draw},
	provector/.style={decorate, decoration={snake,amplitude=2.5pt}, draw},
	antivector/.style={decorate, decoration={snake,amplitude=-2.5pt}, draw},
	fermion/.style={draw=black, postaction={decorate},
		decoration={markings,mark=at position .55 with {\arrow[draw=black]{>}}}},
	fermionbar/.style={draw=black, postaction={decorate},
		decoration={markings,mark=at position .55 with {\arrow[draw=black]{<}}}},
	fermionnoarrow/.style={draw=black},
	gluon/.style={decorate, draw=black,
		decoration={coil,amplitude=4pt, segment length=5pt}},
	scalar/.style={dashed,draw=black, postaction={decorate},
		decoration={markings,mark=at position .55 with {\arrow[draw=black]{>}}}},
	scalarbar/.style={dashed,draw=black, postaction={decorate},
		decoration={markings,mark=at position .55 with {\arrow[draw=black]{<}}}},
	scalarnoarrow/.style={dashed,draw=black},
	electron/.style={draw=black, postaction={decorate},
		decoration={markings,mark=at position .55 with {\arrow[draw=black]{>}}}},
	bigvector/.style={decorate, decoration={snake,amplitude=4pt}, draw},
}
\tikzstyle{block} = [draw, rectangle, minimum height=3em, minimum width=6em]
\usepackage{hyperref}

\usepackage{titling}
\usepackage{subcaption}
\newcommand{\subtitle}[1]{%
	\posttitle{%
		\par\end{center}
	\begin{center}\large#1\end{center}
	\vskip0.5em}%
}
\begin{document}

\begin{center}

\vspace*{15mm}
\vspace{1cm}
{\Large \bf Unitarity Constraints and Collider Searches for Dark Photons}

\vspace{1cm}

{\bf Yasaman Hosseini  and Mojtaba Mohammadi Najafabadi }

 \vspace*{0.5cm}

{\small\sl 
School of Particles and Accelerators, Institute for Research in Fundamental Sciences (IPM) P.O. Box 19395-5531, Tehran, Iran } \\

\vspace*{.2cm}
\end{center}

\vspace*{10mm}

%
%
\begin{abstract}\label{abstract}
Dark photons are predicted by various new physics
models, and are being intensively studied  in a variety of experiments.
In the first part of this paper, we obtain partial wave unitarity constraints on the dark photon parameter space
from the allowed $VV\rightarrow VV$ scattering processes in the limit of large
center-of-mass energy, where $V=W,Z$. 
In the second part of the paper, searches are performed using the expected 
differential rates with a realistic  detector simulation 
including a comprehensive set of background
processes on dilepton and dilepton plus a photon events at the High Luminosity LHC.
In these searches, sensitive differential distributions are used in an optimized way to
determine the sensitivity to dark photon parameter space. 
It is shown that remarkable sensitivity to the
dark photon model is achieved and kinetic mixing strength can be probed down to $(1.4-10)\times 10^{-4}$
for dark photon mass between $15$ GeV to $2$ TeV. 
We also investigate the sensitivity of a future muon collider
suggested by the Muon Accelerator Program (MAP) to the dark photon model at different 
center-of-mass energies. It is shown that 
a future muon collider is able to reach a sensitivity to kinetic mixing at the order of $10^{-4}$.
\end{abstract}

\newpage

\section{Introduction}

Hidden sector states show up in several extensions of the Standard Model (SM) \cite{hs1,hs2,curtin,hs3,hs4,hs5,hs6,hs7} 
to explain thermal relic WIMP dark matter \cite{Pospelov:2007mp, Feldman:2006wd,Feng:2010gw, Porter:2011nv}, 
electroweak baryogenesis \cite{Morrissey:2012db}, hierarchy and naturalness \cite{Craig:2014aea, Craig:2013fga, Chacko:2005pe}.
The hunt for  dark sector physics and hidden sector degrees of freedom are 
one of the major components of the physics program at the High Luminosity LHC (HL-LHC)
and future colliders, such as future circular collider (FCC) \cite{FCC:2018evy, FCC:2018vvp, FCC:2018byv}, 
lepton colliders \cite{Aicheler:2012bya, Fujii:2015jha, Moortgat-Pick:2015lbx}, 
and future muon collider \cite{muc1,muc2,muc3}. 

In general, physics of dark sector could be characterized by the dark sector particle content 
and the mediators that connect the dark states to the SM fields. The interactions between the SM
and dark sector content could be written either in terms of renormalizable or non-renormalizable 
operators \cite{Fabbrichesi:2020wbt}.  The approach of non-renormalizable operators  is suitable 
when the masses of dark sector mediators are larger than the energy scale of the process
under consideration. In such a case, the mediators can be integrated out, and the interactions can be described in
terms of contact non-renormalizable operators. 
On the contrary, if the mediators are light with respect to the energy scale of the process in the 
experiment, they can be produced on-shell and the interactions could be described in
terms of renormalizable operators which is the approach we follow in the present work.
So far, substantial study has been performed to search for dark photon ($Z_{D}$) 
which is a hypothetical massive vector boson mediating the interactions of dark matter
particles \cite{cmsmu,s1,s2,s3,s4,s5,s6,s7,s8,s9,s10,s11,s12, s13}. Dark photon does not directly couple
to SM fields, however, it can receive a small coupling to the electromagnetic current from
the kinetic mixing between the $Z_{D}$ and SM hypercharge. 
This coupling, which is tiny with respect to that of the SM photon by a factor labeled $\epsilon$, would
provide a possibility through which $Z_{D}$ can be produced in the laboratory, and also
it allows  $Z_{D}$ to decay into visible SM particles \cite{Fabbrichesi:2020wbt}.
The kinetic mixing $\epsilon$ is an arbitrary parameter, however, special ranges of $\epsilon$
are interesting which come from the loop level effects of heavier particles .
Particularly, quantum effects of a heavy state that carries both SM hypercharge and $U(1)_{D}$
charge typically generate $\epsilon \sim 10^{-4}-10^{-2}$ \cite{Holdom:1985ag,Arkani-Hamed:2008kxc,Baumgart:2009tn}.
Therefore, probing $\epsilon$ in the motivated region is one of the main goals of 
searches in dark sector physics.
The value of kinetic mixing $\epsilon$ determines the dark photon lifetime.
For small kinetic mixing values, $Z_{D}$ has a long lifetime therefore it decays at a macroscopic distance
from the point it is produced \cite{curtin}. In this work the concentration is on  values of the kinetic mixing  
with $m_{Z_{D}}$  above $\sim 20$ GeV  that  the dark photons decay promptly.

Clarification of the open questions in the SM and the observed
consistency between the current LHC data and SM predictions has increased 
the desire to HL-LHC and has revived the interest
of the community into a multi-TeV muon collider program \cite{muc1,muc2,muc3}. 
One of the marvelous advantages of utilising muon beams  is less radiation than electrons in a circular collider.
While there are challenges such as the difficulties of achieving high intensity
and low emittance muon beams, recent developments as well as potential alternative ways
to achieve high intensity muon beams may lead to overcome the present limitations
in the coming few years. Therefore, multi-TeV muon colliders could provide a new way
to reach the energy frontier and to allow  direct production of new
heavy states. 

In the first part of this paper, we derive constraints on dark photon interactions from partial wave
unitarity, examining the allowed $VV\rightarrow VV$ scattering processes in the limit of
large center-of-mass energies, where $V=W,Z$. 
The second part of the article is dedicated to collider searches for the dark sector. 
Dilepton production and dilepton production in association with a photon at the HL-LHC are
revisited as processes to probe dark photon. Drell-Yan process at the HL-LHC has been already used
in Refs. \cite{curtin, Hoenig:2014dsa, Cline:2014dwa} in search for dark photon. In the present work, instead of repeating their analyses
a more detailed study using the expected differential rates including realistic detector effects 
and all sources of background processes is performed.
As a complementary channel, a study is also carried out using 
dilepton production plus a photon at the HL-LHC. 
For this process, a test statistic is performed on the invariant mass of final state to derive  
sensitivity. 
Then, for the first time, Drell-Yan process at a multi-TeV muon 
collider is used to probe the dark sector.  Due to the high energy and precision, the future multi-TeV
muon colliders offer a golden opportunity to probe dark photon parameter space.

The organization of this paper is as follows. Section \ref{sec:th} briefly describes the theoretical framework.
 Section \ref{sec:vvvv} is dedicated to obtain the unitarity constraints using the $VV\rightarrow VV$
 scatterings at high energies.
In Section \ref{sec:col}, we describe the analyses for 
determining the potential of HL-LHC to probe dark photons using the Drell-Yan and 
Drell-Yan in association with a photon in proton-proton collisions at $\sqrt{s} = 14$ TeV
using 3000 fb$^{-1}$ integrated luminosity of data.
In section \ref{sec:col}, we also obtain prospects of a future multi-TeV muon collider
 at different center-of-mass energies and discuss the impact this has on our limit projections. Section \ref{sec:conclusions} contains
our conclusions.

\section{Theoretical framework}
\label{sec:th}

Dark sectors commonly comprise one or more mediator states which couple to the SM through a
portal. The portal describing the interactions of dark sector with SM is dependent on the spin and parity of  mediators.
In general, the mediator can be a fermion, a vector, a scalar, or a pseudoscalar. 
In this work, we restrict ourselves to a simple dark photon model based on the extension of the SM
gauge group to $\rm SU(3)_{C}\times SU(2)_{L}\times SU(2)_{Y} \times U(1)_{D}$, where there is 
a kinetic mixing between a dark Abelian gauge symmetry, $\rm U(1)_{D}$ and the SM $\rm U(1)_{Y}$. In such an approach, a dark Higgs
field $S$ with non-zero vacuum expectation value is used to break the new dark $\rm U(1)_{D}$ symmetry. The
related gauge terms in the Lagrangian are:
\begin{eqnarray}
\mathcal{L}  \supset -\frac{1}{4}\hat{Z}_{D\mu\nu}\hat{Z}^{\mu\nu}_{D}-\frac{1}{4}\hat{B}_{\mu\nu}\hat{B}^{\mu\nu}
+\frac{1}{2}\frac{\epsilon}{\cos\theta_{W}}\hat{Z}^{\mu\nu}_{D}\hat{B}_{\mu\nu},
\end{eqnarray}
where the hatted fields are gauge eigenstates, 
$\hat{Z}^{\mu\nu}_{D} = \partial^{\mu}\hat{Z}^{\nu}_{D} - \partial^{\nu}\hat{Z}^{\mu}_{D}$ and 
$\hat{B}^{\mu\nu}= \partial^{\mu}\hat{B}^{\nu} - \partial^{\nu}\hat{B}^{\mu}$ 
 denote the dark photon and SM hypercharge field strengths,
respectively.  $\cos\theta_{W}$ is the cosine of Weinberg angle and 
 $\epsilon$ is the kinetic mixing parameter.
The mass eigenstates of gauge boson are obtained by performing a rotation on the three neutral
components of the gauge fields $\hat{B}, \hat{Z}_{D}$, and $W_{3}$ where a new mixing angle $\alpha$
appears:
\begin{eqnarray}
\tan\alpha = \frac{1-\delta^{2}-\eta^{2}\sin^{2}\theta_{W}- {\rm Sign}(1-\delta^{2})\sqrt{4\eta^{2}\sin^{2}\theta_{W}+(1-\eta^{2}\sin^{2}\theta_{W}-\delta^{2})^{2} } }{2\eta\sin\theta_{W}},
\end{eqnarray}
where 
\begin{eqnarray}
\eta = \frac{\epsilon}{\cos\theta_{W}\sqrt{1-\frac{\epsilon^{2}}{\cos^{2}\theta_{W}}}},
\end{eqnarray}
and $\delta$ relates the dark vector mass to SM $Z$-boson mass before mixing, i.e. 
$m^{2}_{Z_{D},0}  \equiv \delta^{2} \times m^{2}_{Z,0}$ where $m_{Z_{D},0}$ and $m_{Z,0}$
are the dark photon and SM $Z$-boson masses before mixing. $\theta_{W}$ is the Weinberg angle.

More details 
of the formulating of the model can be found in Refs.\cite{curtin, Fabbrichesi:2020wbt}.
The interaction between $Z,Z_{D}$-boson and the SM fermions,  has a form of $g_{Z_{(D)}f\bar{f}}Z_{\mu(D)}\bar{f}\gamma^{\mu}f$, where 
\begin{eqnarray}
g_{Zf\bar{f}} =  \frac{g}{\cos\theta_{W}} (\cos\alpha(T^{3}\cos^{2}\theta_{W} - Y\sin^{2}\theta_{W}) + \eta\sin\alpha\sin\theta_{W}Y), \nonumber \\
g_{Z_{D}f\bar{f}} = \frac{g}{\cos\theta_{W}}(-\sin\alpha(T^{3}\cos^{2}\theta_{W} - Y\sin^{2}\theta_{W}) + \eta\cos\alpha\sin\theta_{W} Y),
\end{eqnarray}
where $f$ is the left- or right-handed fermion, $Y$ and $T^{3}$ are the third component
of hypercharge  and weak isospin of the fermion $f$, respectively. 
 The potential describing the SM Higgs and the dark Higgs sector has the following form:
\begin{eqnarray}
V(H,S) = -\mu^{2}|H|^{2}+\lambda|H|^{4} -\mu^{2}_{S}|S|^{2}+\lambda_{S}|S|^{4}+\kappa_{HS}|H|^{2}|S|^{2}. 
\label{pot}
\end{eqnarray}
where $S$ is the singlet dark Higgs field and the SM Higgs doublet is denoted by $H$.
The dark Higgs $S$ and SM Higgs fields are linked through a renormalizable parameter $\kappa_{HS}$.
Requiring the stability of the potential at large field values 
points to the requirement $ \lambda, \lambda_{S} > 0$ and $\kappa_{HS} \ge -2\sqrt{\lambda\lambda_{S}}$ \cite{xx}. 
The constraint obtained from the requirement of  perturbative unitarity
leads to $\kappa_{HS} \leq 8\pi$ \cite{cx}.
Indirect and direct constraints on dark Higgs mass and the corresponding mixing 
have been obtained in Ref.\cite{adam}.
After spontaneous symmetry breaking, the mixing between the dark Higgs and SM Higgs is 
defined by $\theta_{h}$ angle which is related to $\kappa_{HS}$ in Eq. \ref{pot} via the following relation:
\begin{eqnarray}
\tan\theta_{h} = \frac{\lambda v^{2}-\lambda_{S}v_{S}^{2}-\text{Sign}(\lambda v^{2}-\lambda_{S}v_{S}^{2})\sqrt{\lambda^{2}v^{4}+
\lambda_{S}^{2}v_{S}^{4}+v^{2}v_{S}^{2}(\kappa_{HS}^{2}-2\lambda\lambda_{S})}}{\kappa_{HS} vv_{S}}.
\end{eqnarray}
where $v$ and $v_{S}$ are the vacuum expectation values of the SM Higgs and the singlet scalar
which generates dark Higgs mass. For small mixing angles \cite{curtin}:
\begin{eqnarray}
\tan\theta_{h}  \approx  \frac{\kappa_{HS}}{2}\times\frac{vv_{S}}{\lambda_{S}v_{S}^{2} - \lambda v^{2}}.
\end{eqnarray}


There are several extensive experimental and phenomenological studies on 
searching for  dark photon. The LHCb experiment searches are done
for both prompt-like and long-lived dark photons using 
the events produced in proton-proton collisions at a center-of-mass energy of 13 TeV \cite{s3}. The
search is based on the dimuon decays using data  corresponding to an
integrated luminosity of 5.5 fb$^{-1}$. No 
evidence for dark photon has been found, and stringent exclusion limits at $90\%$ confidence level  are determined
on the kinetic mixing strength on the mass region $0.214 < m_{Z_{D}}  < 0.740$  GeV and $10.6  < m_{Z_{D}} < 30$ GeV. 
Another search has been performed by the CMS collaboration for a narrow resonance decaying to a dimuon pair 
using proton-proton collision data corresponding to an integrated luminosity of 137 fb$^{-1}$ \cite{cmsmu}. 
No deviation from the SM prediction has been observed in the explored mass ranges. 
Constraints on the kinetic mixing have been derived over the mass region of dark photon in the range 
of $30-75$ GeV and $110-200$ GeV.

In Ref.\cite{curtin}, it has been shown that the direct Drell-Yan production provides
high sensitivity to dark photon, and can probe $\epsilon \geq 9 \times 10^{-4}$ at the 
HL-LHC and $ \epsilon \geq 4\times 10^{-4}$ at the FCC-hh with $\sqrt{s} = 100$ TeV.
Indirect constraints from the global fits to electroweak precision
observables from the measurements at LEP,
Tevatron, and the LHC have been also obtained in Ref.\cite{curtin} which excludes 
$\epsilon \gtrsim 3 \times 10^{-2}$
for $5 \lesssim m_{Z_{D}} \lesssim 100$ GeV. 
There are  bounds on  dark photon parameters from beam-dump experiments \cite{bd1,bd2,bd3}, 
fixed-target experiments \cite{ft}, and rare meson decays \cite{md}.
In Ref.\cite{DOnofrio:2019dcp},  an estimate for  a dark
photon search at the LHeC and FCC-he through the dark photon displaced
decays into two charged particles and in a mass range of
10 to 700 MeV has been studied.  The LHeC (FCC-he)
can exclude dark photons in the considered mass range with
$\epsilon$ larger than $2\times10^{-5}(10^{-5})$.

In the next section, limits on dark photon interactions from partial wave
unitarity are derived from the $VV\rightarrow VV$ scattering processes.

\section{Unitarity constraints}
 \label{sec:vvvv}

Violation of the partial wave unitarity  indicates that new fundamental degrees of freedom or new composite states,
must be present around or below the scale of unitarity violation to preserve a physical behaviour of scattering amplitudes.
Unitarity requirement imposes consistency conditions on the theory parameters to be valid up to a given energy scale. 
On the other hand, for given values of the dark photon parameters, unitarity imposes an upper bound on the energy scale at which the 
theory is valid. We obtain constraints on the kinetic mixing from the partial wave
unitarity, examining  $VV \rightarrow VV$ scattering processes in the limit of large center-of-mass energy.  
A natural version of the unitarity requirement is \cite{u1}:
\begin{eqnarray}
  |\Re(a_0)| \leq \dfrac{1}{2},
         \label{azero}
\end{eqnarray}
where $a_{0}$ is corresponding to the $J=0$ partial wave: 
\begin{equation}
   a_0 = \dfrac{1}{32 \pi } \int_{-1}^{+1} \mathcal{M}(\cos\theta) d\cos\theta.
\end{equation}

\begin{figure}{}
   	   \centering
	      		\includegraphics[width=0.5\linewidth]{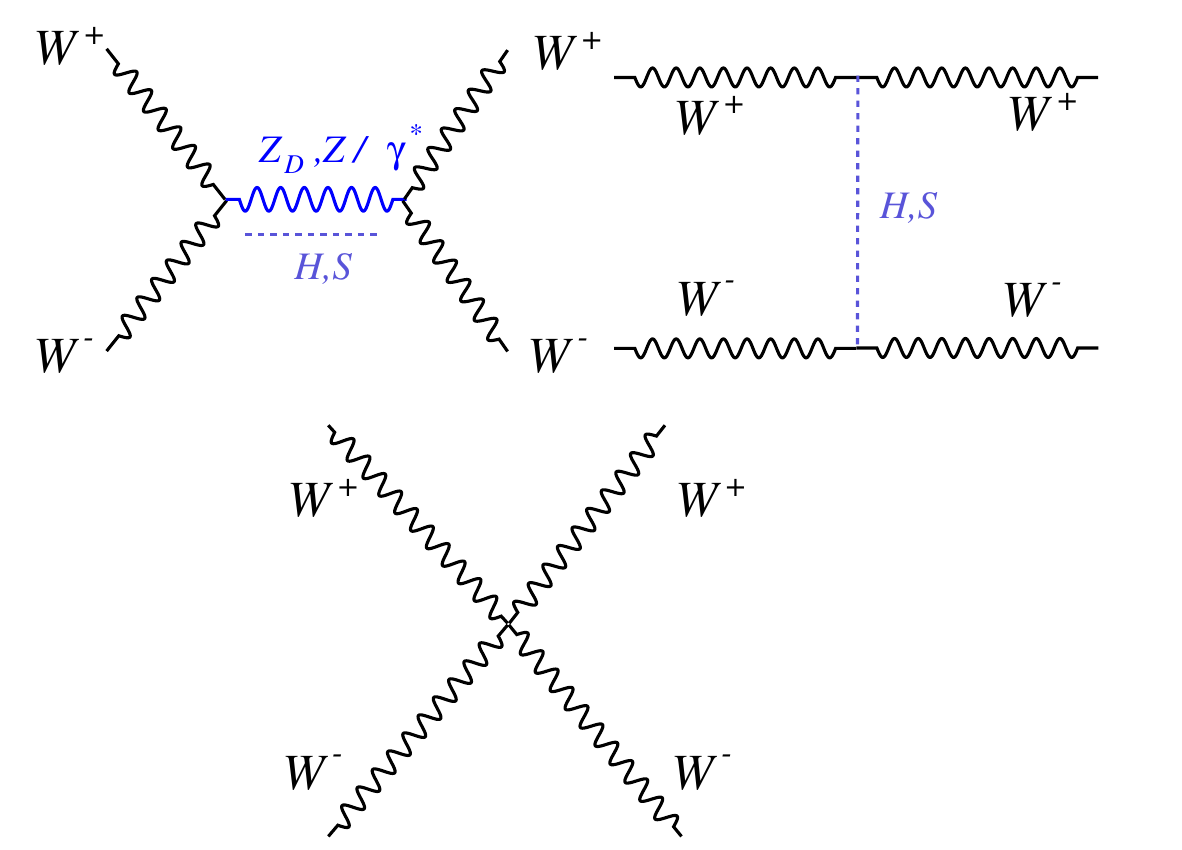}
   	       \caption{Representative Feynman diagrams contributing to $W^{+}W^{-}\rightarrow W^{+}W^{-}$ scattering. In this figure $H$ is the
	       SM Higgs boson and $Z_{D}$ is dark photon.}  
   	       \vspace*{1cm}
	       \label{fig:feynman}
  \end{figure}

 There are several $s$- and $t$-channel diagrams containing the 
 exchange of a dark photon $Z_{D}$, $Z$, $\gamma$, and Higgs boson combined with 
 the four-point interaction for the  elastic scattering of $W^{+}W^{-}\rightarrow W^{+}W^{-}$.
Representative Feynman diagrams for  $W^{+}W^{-}\rightarrow W^{+}W^{-}$ scattering are depicted  in Figure \ref{fig:feynman}.
In addition to $Z_{D}$, there are contributions from dark Higgs boson to the $W^{+}W^{-}\rightarrow W^{+}W^{-}$ scattering.
These contributions are expected to be small in low mass region of dark Higgs boson as the coupling of dark Higgs with $W^{+}W^{-}$ is proportional to
$ m_{W}^{2}\sin\theta_{h}/v$ \cite{adam}. At low mass region $m_{S} \lesssim 10$ GeV, any value of $\theta_{h}$ above $\sim 10^{-4}$ has been excluded from 
LHCb \cite{e1,e2}, BNL-E949 \cite{e3}, CHARM \cite{e4}, LSND \cite{e5}, and MicroBooNE \cite{e6} experiments.
A summary of the reach from the present and proposed experiments for a dark Higgs boson are depicted in Ref.\cite{ccvv}.
As a result,  the contribution of light dark Higgs in the partial wave amplitude $a_{0}$ is  suppressed by a factor of 
$(m_{W}^{2}\sin\theta_{h}/v)^{2} \lesssim \mathcal{O}(10^{-6})$.

\begin{figure}[t]
   	   \centering
	      		\includegraphics[width=0.3\linewidth]{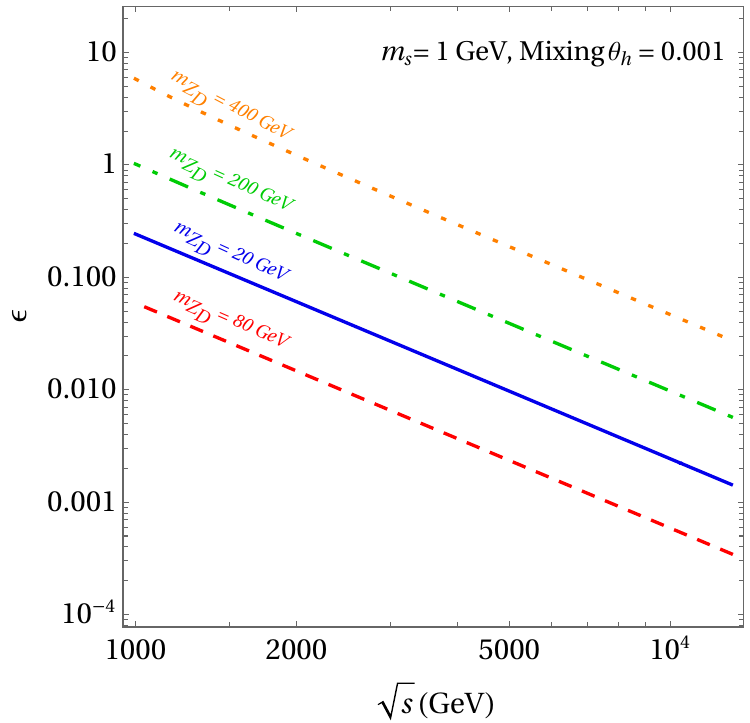}
   		   \includegraphics[width=0.3\linewidth]{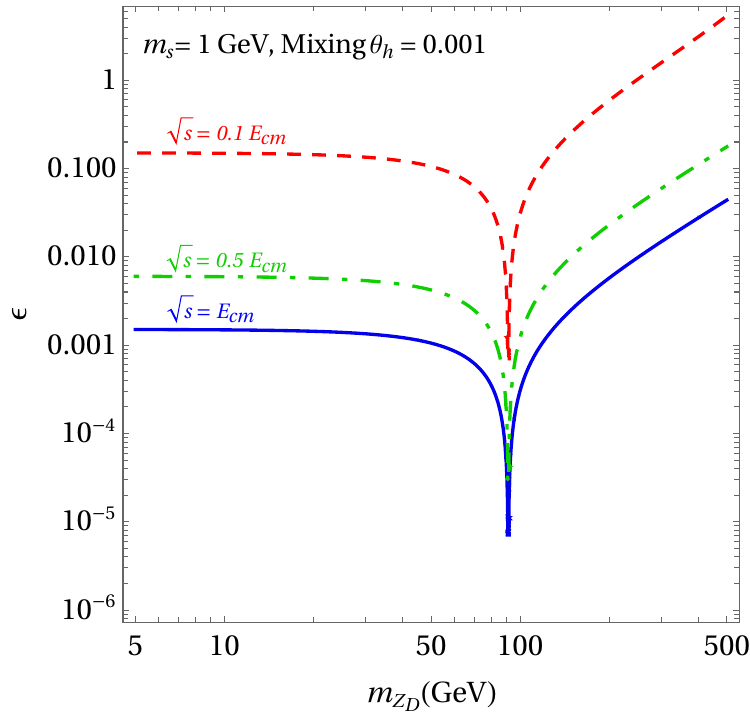}\\
		  \includegraphics[width=0.3\linewidth]{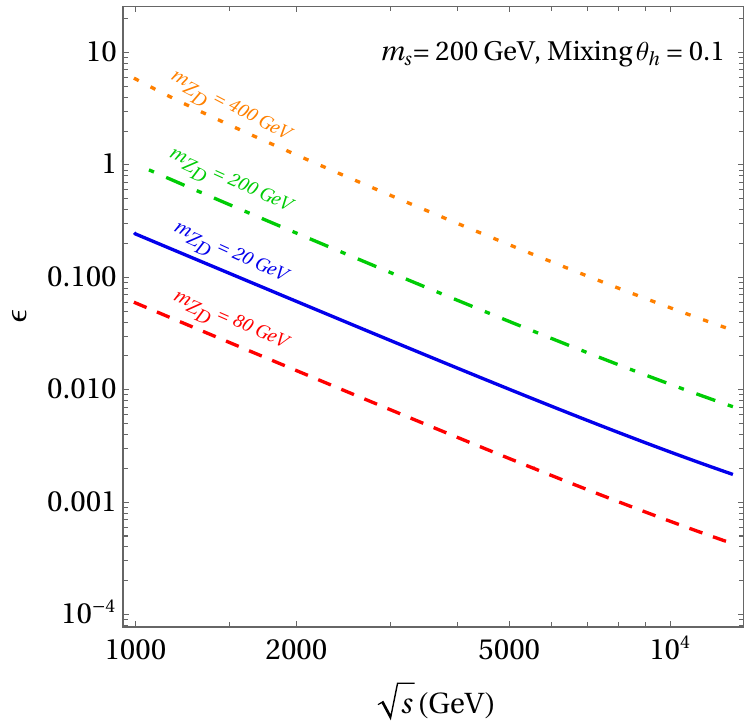}
   		   \includegraphics[width=0.3\linewidth]{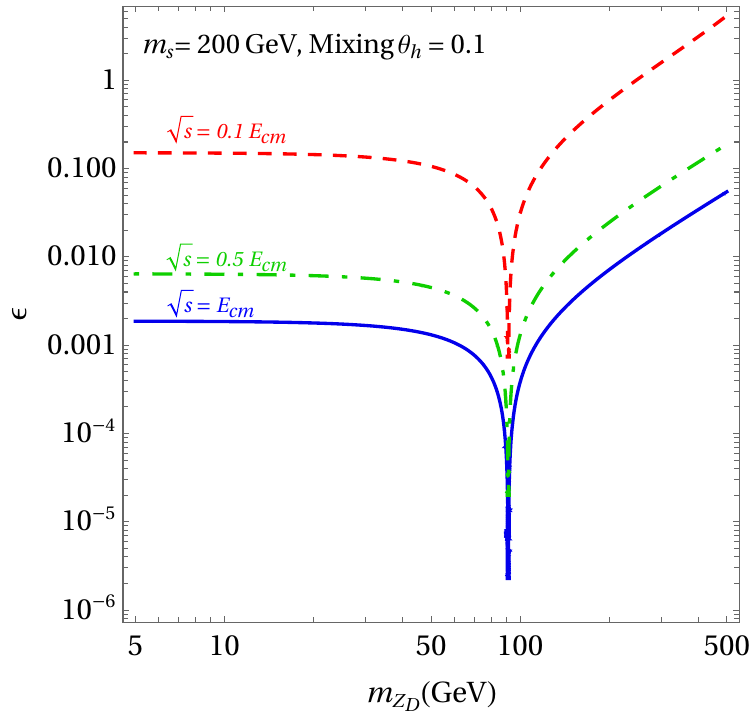}
   	       \caption{Left: the upper limit on $\epsilon$ is shown versus $\sqrt{s}$
               for $m_{Z_{D}} = 20$, 80, 200, and 400 GeV assuming $(\theta_{h} = 0.001, m_{S} = 1 ~\text{GeV})$ (top)
                and $(\theta_{h} = 0.1, m_{S} = 200 ~\text{GeV})$ (bottom). 
               Right: upper bound on $\epsilon$ in terms of $m_{Z_{D}}$ for three choices 
               of center-of-mass energies of $\sqrt{s} = 0.1E_{cm}, 0.5E_{cm}$ and $E_{cm}$, where $E_{cm}$ is assumed
               to be 13 TeV. The bounds are shown for $(\theta_{h} = 0.001, m_{S} = 1~ \text{GeV})$ (top)
                and $(\theta_{h} = 0.1, m_{S} = 200~ \text{GeV})$ (bottom).}  
	       \label{fig:limit}
  \end{figure}

Total amplitude for all contributing diagrams in the high energy limit has the following form:
   \begin{eqnarray}
\begin{split}
\mathcal{M}_{total} &= -\dfrac{g^2 m_H^2}{4 m_W^2} \left( \dfrac{s}{s - m_H^2} + \dfrac{\dfrac{s}{2} \left(1 - \cos \theta\right)}{\dfrac{s}{2} \left(1 - \cos \theta\right) + m_H^2}  \right) - \dfrac{\sin^2 \theta_h s}{2 v^2} \left(1+ \cos \theta\right) \\
&- \dfrac{\sin^2 \theta_h m_S^2}{v^2} \left( \dfrac{s}{s - m_S^2} + \dfrac{\dfrac{s}{2} (1- \cos \theta)}{\dfrac{s}{2} (1- \cos \theta) + m_S^2}\right) - \dfrac{g^2 s}{4 m_W^4} \sin^2\alpha \cos^2\theta_W \\
&\times \left(\dfrac{3 s}{2} \cos \theta + \dfrac{3 m^2_{Z_D}}{2} \left(1 + \cos \theta\right) + \dfrac{s}{4}\left(\cos^2 \theta - 3\right) - 8 m_W^2 \cos \theta \right),
\end{split}       \label{M_total}
   \end{eqnarray}
where $s$ is the center-of-mass energy, $\theta$ is the scattering angle,
$m_{H}$ is SM Higgs boson mass, $m_S$ is dark Higgs mass, and $m_{W}$ is $W$ boson mass.
After integration over $\cos\theta$ from $-1$ to $1$, we find:
   \begin{equation}
\begin{split}
\int_{-1}^{+1} \mathcal{M}_{total} d\cos \theta &= -\dfrac{g^2 m_H^2}{4 m_W^2} \left(\dfrac{2 s}{s - m_H^2} + 2 - \dfrac{2 m_H^2}{s} \ln\left[1 + \dfrac{s}{m_H^2}\right]\right)\\
&- \dfrac{\sin^2 \theta_h s}{v^2} - \dfrac{\sin^2 \theta_h m_S^2}{v^2} \left(\dfrac{2s}{s - m_S^2} + 2 - \dfrac{2 m_S^2}{s} \ln\left[1 + \dfrac{s}{m_S^2}\right]\right)\\
&- \dfrac{g^2 s}{4 m_W^4} \sin^2 \alpha \cos^2 \theta_W \left(3 m_{Z_D}^2 - \dfrac{4}{3} s\right).
\end{split}       \label{M_total_integrated}
   \end{equation}
 As it can be seen the scattering amplitude grows as the center-of-mass energy increases. It is notable that the term 
proportional to $s$ arises from the diagram where dark photon $Z_{D}$ is exchanged. 
The contribution of dark photon leads to unitarity
violation at high energies. 
The partial wave amplitude $a_{0}$ in the high energy limit where $m^{2}_{S}, m^2_H \ll s$ becomes:
   \begin{equation}
a_0 = - \dfrac{m_H^2}{8 \pi v^2} - \dfrac{s}{32 \pi v^2 m_W^2} \sin^2\alpha \cos^2\theta_W \left(3 m_{Z_D}^2 - \dfrac{4}{3} s\right) - \dfrac{\sin^2 \theta_h m_S^2}{8 \pi v^2} -
 \dfrac{\sin^2 \theta_h s}{32 \pi v^2}.
       \label{a_zero}
 \end{equation}
 Applying the partial-wave unitarity condition in Eq.\ref{azero}, one finds upper limits on 
 $\epsilon$ for different masses of the dark photon.  In the left panel of Figure \ref{fig:limit},
 the upper limit on kinetic mixing strength $\epsilon$ versus the center-of-mass energy
 for four choices of the dark photon mass and for two cases of 
 $(\theta_{h} = 0.001, m_{S} = 1 ~\text{GeV})$ (top)
 and $(\theta_{h} = 0.1, m_{S} = 200 ~\text{GeV})$ (bottom) is presented.  As expected, the constraint on $\epsilon$ gets tighter
 as $\sqrt{s}$ grows. Among different masses, for $m_{Z_{D}}$ close to the $Z$ boson mass, the best sensitivity
 is achieved however with increasing $m_{Z_{D}}$ weaker bounds on $\epsilon$ is obtained.
 In the right plot of Figure \ref{fig:limit}, upper constraint on $\epsilon$ in terms of dark photon mass for
 three assumptions of center-of-mass energies of $\sqrt{s} = 0.1E_{cm}, 0.5E_{cm}$ and $E_{cm}$ is presented.
 In deriving the limits, the value of $E_{cm}$ is set to 13 TeV. The higher $\sqrt{s}$ the stronger limit on $\epsilon$. 
 We find that for $m_{Z_{D}} \lesssim 50$ GeV, with the center-of-mass energy
 of $\sqrt{s} = 13$ TeV, any value of $\epsilon$ above $10^{-3}$ is excluded by the unitarity conditions.
From a comparison of top and bottom plots in Fig.\ref{fig:limit}, no significant dependence on the dark Higgs and the mixing angle is
 observed which is expected in particular at low mass values. 
 At the end of this section, it is worth mentioning that the bounds derived from $ZZ\rightarrow ZZ$
 are found to be two orders of magnitude looser than those obtained from  $WW\rightarrow WW$ and 
 at lowest order, $Z_{D}Z_{D}\rightarrow Z_{D}Z_{D}$ process provides no constraints on $\epsilon$ since the amplitude 
 has no dependence on $\epsilon$.

\section{ Collider searches }
\label{sec:col}

The planned HL-LHC at 14 TeV with an integrated luminosity of 3000 fb$^{-1}$,
future hadron and electron-positron colliders, and multi-TeV muon colliders provide excellent opportunities to explore dark photon.
In this section, we revisit the potential of the HL-LHC to probe the dark photon using Drell-Yan and Drell-Yan associated with a photon. 
We also present the prospect of a multi-TeV future muon collider to probe dark photon
through Drell-Yan process.  In all of the analyses presented in this section $\epsilon$ and $m_{Z_{D}}$ are probed and
it is assumed that the dark Higgs is heavy and the mixing $\kappa_{HS}$ is set to a very small value ($\kappa_{HS} \sim 10^{-10})$.

\subsection{Dilepton production through dark photon at the LHC}
\label{sec:dy}

One of the processes  which provides an excellent sensitivity to 
dark photon is the Drell-Yan production process,  $pp\rightarrow Z_{D}$.
Among the decay modes of $Z_{D}$, the hadronic decay mode of $Z_{D}$ is the major decay channel, 
however due to large contribution of multijet background it is hard to achieve good sensitivity.
On the other hand,  the leptonic decay mode of $Z_{D}$ has smaller background and
lepton reconstruction and identification efficiency and energy resolution are
much better than those for jets.
In Ref. \cite{curtin, Hoenig:2014dsa, Cline:2014dwa}, the HL-LHC potential to search for $Z_{D}$ have been estimated 
using the Drell-Yan process in the presence of dark photon. 
In this section, it is shown that the $|\Delta\eta| = |\eta_{\ell^{+}} - \eta_{\ell^{-}}|$ distribution is
a sensitive variable to search for dark photon in $pp\rightarrow Z_{D} \rightarrow \ell^{+}+\ell^{-}$.
We consider the main sources of background processes and take into account a CMS-like 
detector effects to probe the parameter space of dark photon.
The dominant SM backgrounds to the assumed signal process are: 
SM Drell-Yan $pp\rightarrow Z/\gamma^{*} \rightarrow \ell^{+}+\ell^{-}$; production of two massive gauge bosons $WW,WZ,ZZ$
where the $W$ and $Z$ bosons can decay  leptonically and/or hadronically;  Drell-Yan production of $\tau^{+}\tau^{-}$ with their 
subsequent decays to electron and muon pairs; $t\bar{t}$ in particular in 
the dileptonic decay mode; single top in $tW$-channel; and misidentified contribution where jets are misidentified 
as leptons. 

\begin{figure}
   \centering
   	   \includegraphics[width=0.5\linewidth]{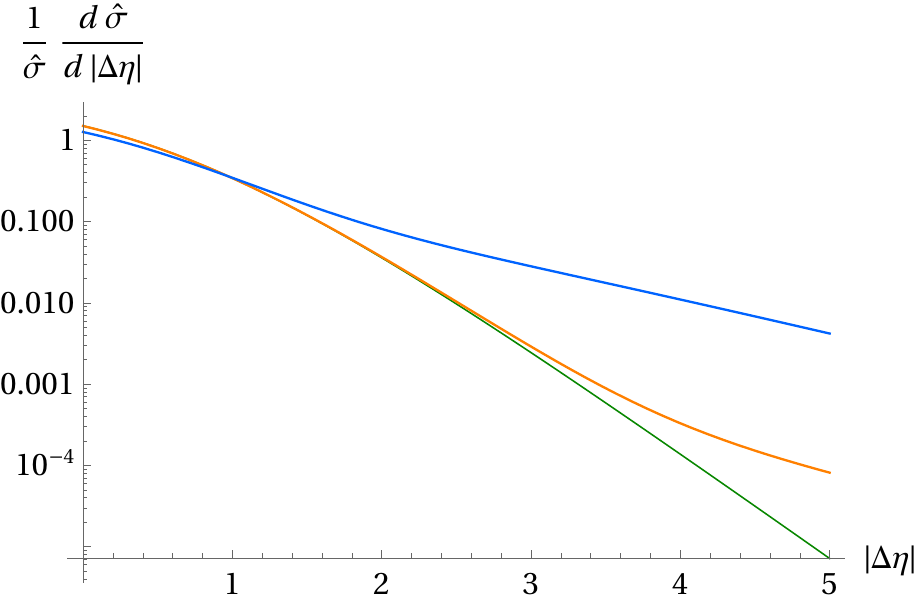}
	   \caption{\small The differential partonic cross section of $|\Delta\eta| = |\eta_{\ell^{+}}-\eta_{\ell^{-}}|$
	    for the SM,  $m_{Z_{D}} = 50$ and  $1200$ GeV assuming $\epsilon$ = 0.05. The SM expectation is depicted by the green curve and the orange
	    and blue curves present the SM+$Z_{D}$ with $m_{Z_{D}} = 50$ and  $1200$ GeV, respectively.}
	   \label{DY_sigma_Deta}
   \end{figure}

Consider the lepton pair production through an intermediate $Z_{D},Z$ boson
or a photon in proton-proton collisions at the LHC. The cross section and differential cross sections can be 
factorized as follows:
\begin{eqnarray}
\frac{d\sigma(pp\rightarrow \ell^{+}\ell^{-})}{dm_{\ell\ell}} = \sum_{q,\bar{q}}\int dx_{1}dx_{2}f_{q}(x_{1},Q)f_{\bar{q}}(x_{2},Q)\times \frac{d\hat{\sigma}(q\bar{q}\rightarrow \ell^{+}\ell^{-})}{dm_{\ell\ell}},
\label{eq:sigma}
\end{eqnarray}
where $m_{\ell\ell}$ is the invariant mass of dilepton, $x_{1,2}$
are the momentum fractions of the quark $q$ and anti-quark $\bar{q}$
partons in the protons and $f_{q,\bar{q}}(x_{i},Q)$ are the proton parton distribution functions which are dependent on a factorization scale $Q$. 
The invariant mass is related to the center-of-mass energy via $m_{\ell\ell} = \sqrt{\hat{s}} = \sqrt{x_{1}x_{2}S}$, where $\sqrt{\hat{s}}$ and  $\sqrt{S}$
are the partonic center-of-mass energy and center-of-mass energy, respectively. 
The cross section of dilepton production consists of terms arising from exchange of $\gamma, Z, Z_{D}$, as well as the interference terms \cite{zp}:
\begin{eqnarray}
 \dfrac{d\sigma (pp\rightarrow \ell^{+}\ell^{-})}{dm_{\ell\ell}} = \dfrac{d\sigma_{\gamma\gamma}}{dm_{\ell\ell}} + \dfrac{d\sigma_{ZZ}}{dm_{\ell\ell}} + \dfrac{d\sigma_{Z_{D} Z_{D}}}{dm_{\ell\ell}}
 + 2 \dfrac{d\sigma_{\gamma Z}}{dm_{\ell\ell}}  + 2 \dfrac{d\sigma_{ \gamma Z_{D}}}{dm_{\ell\ell}} 
 + 2 \dfrac{d\sigma_{ZZ_{D}}}{dm_{\ell\ell}}. \label{sigma_eq}
 \end{eqnarray}
The differential  cross section of pure signal reads
\begin{eqnarray}
 \dfrac{d\sigma_{Z_{D} Z_{D}}}{dm_{\ell\ell}} = \sum_{q,\bar{q}} d\eta_{\ell^{+}} d\eta_{\ell^{-}} x_1 f_q(x_1) x_2 f_{\bar{q}}(x_2) \dfrac{2}{\sqrt{\hat{s}}}
  \dfrac{d \hat{\sigma}_{Z_{D} Z_{D}}}{d\Delta\eta},
\end{eqnarray}
where
  \begin{eqnarray}
  \dfrac{d\hat{\sigma}_{Z_{D} Z_{D}}}{d\Delta\eta} = \dfrac{1}{32 \pi N_c \cosh^2\frac{\Delta\eta}{2}} \dfrac{\hat{s}}{(\hat{s} - m^2_{Z_{D}})^2 + m^2_{Z_{D}} \Gamma^2_{Z_{D}}} \sum_{i = 0}^{2} c_i^q \left(-\dfrac{e^{-\frac{\Delta\eta}{2}}}{2 \cosh \frac{\Delta\eta}{2}}\right)^i,
   \end{eqnarray}
where $N_{c} = 3$, $\Gamma_{Z_{D}}$ is the total width of dark photon.
The partonic differential cross section for the interference terms has the following form:
\begin{equation}
       \dfrac{d\hat{\sigma}_{ij}}{d{\Delta\eta}} = \dfrac{\hat{s}}{32 \pi N_c} \dfrac{(\hat{s} - m_i^2)(\hat{s} - m_j^2) + 
       m_i m_j \Gamma_i \Gamma_j}{\left[\left(\hat{s} - m^2_{i}\right)^2 + m^2_{i} \Gamma^2_{i}\right] \left[\left(\hat{s} -
        m^2_{j}\right)^2 + m^2_{j} \Gamma^2_{j}\right]} c_{2,ij}^q \dfrac{1}{\cosh^2 \frac{\Delta\eta}{2}} \left(-\dfrac{e^{-\frac{\Delta\eta}{2}}}{2 \cosh \frac{\Delta\eta}{2}}\right)^2 ,
 \end{equation}
where $i,j = \gamma, Z, Z_{D}$ and the $c_{0,1,2}^{q}$ coefficients are defined as:
 \begin{eqnarray}
       c_0^q &=& \left[(g^V_q)^2 + (g^A_q)^2 \right]. \left[(g^V_\ell)^2 + (g^A_\ell)^2 \right] - 4 g^V_q g^A_q g^V_{\ell} g^A_{\ell}, ~c_1^q = 2 c_0^q, \nonumber \\
       c_2^q &=& \left[(g^V_q)^2 + (g^A_q)^2 \right]. \left[(g^V_\ell)^2 + (g^A_\ell)^2 \right], \\
       c_{2,ij}^q &=& 2 \left[g^V_{q,i} g^V_{q,j} + g^A_{q,j} g^A_{q,j}\right].\left[g^V_{\ell,i} g^V_{\ell,j} + g^A_{\ell,j} g^A_{\ell,j}\right], \nonumber
   \end{eqnarray}
The vector and axial couplings $g^{V,A}_{i}$ for leptons and quarks to $\gamma,Z$, and $Z_{D}$ are:
 \begin{eqnarray}
\gamma:  g_{f}^{V} &=& e Q_{f}, \quad g_{f}^{A} = 0, \nonumber \\
Z:  g_f^V &=& \dfrac{e}{\sin\theta_W \cos\theta_W} \left(\cos\alpha(\dfrac{T_3}{2} - Q_{f} \sin^2\theta_W) - \eta \sin\alpha \sin\theta_W (\frac{T_{3}}{2}-Q_{f})\right), \nonumber \\
Z:  g_f^A &=& \dfrac{ e}{\sin\theta_W \cos\theta_W} \left( \cos\alpha - \eta\sin\alpha\sin\theta_{W}\right) \dfrac{T_3}{2},   \\
Z_{D}: g_f^V &=& \dfrac{e}{\sin\theta_W \cos\theta_W} \left(- \sin\alpha \left(\frac{T_3}{2}  - Q_{f} \sin^2\theta_W\right) - \eta \cos\alpha \sin\theta_W (\frac{T_{3}}{2}-Q_{f}) \right),  \nonumber\\
Z_{D}: g_f^A &=& \dfrac{ e}{\sin\theta_W \cos\theta_W} \left(- \sin\alpha - \eta\cos\alpha\sin\theta_{W}\right)\frac{T_{3}}{2} \nonumber. 
\end{eqnarray}
where $Q_{f}$ is fermion electric charge.
Because the main irreducible background, i.e. Drell-Yan, has exactly the same initial and final states as
signal, there is interference between signal and background at tree level appeared in Eq.\ref{sigma_eq}
by $\sigma_{\gamma Z_{D}}$ and $\sigma_{Z Z_{D}}$.  In Fig. \ref{DY_sigma_Deta}, the
normalized differential partonic cross section for signal+background and for Drell-Yan only is depicted.
The signal curves belong to two choices of $m_{Z_{D}} = 50$ and $1200$ GeV with $\epsilon = 0.05$. 
 As one can see, at large $\Delta\eta$ values, the dilepton production rate gets separated from 
 the SM Drell-Yan process and goes up visibly above the SM. 
This discrepancy is expected to be more important at  large $\hat{s}$ 
 which is corresponding to large $m_{Z_{D}}$. Therefore, more sensitivity to heavy dark photons is expected.
 In the following, this difference in shape is used 
 to probe the dark photon parameter space. 
      
\begin{figure}{}
     	   \centering
	      	\includegraphics[width=0.5\linewidth]{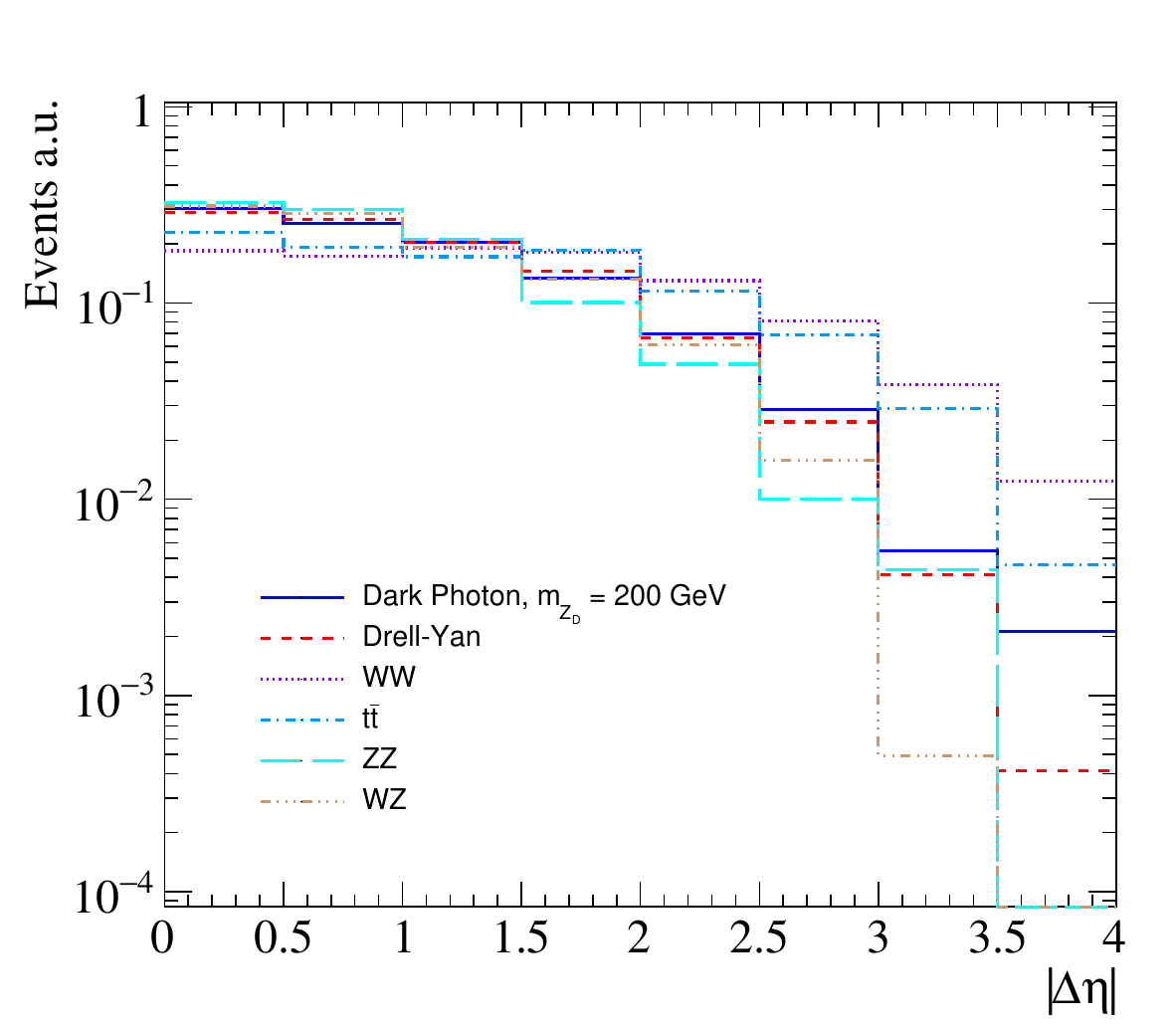}
   	       \caption{ \small The normalized distribution of $|\Delta \eta| = |\eta_{\ell^{+}}-\eta_{\ell^{-}}|$ for dark photon signal with $m_{Z_{D}} = 200$ GeV
	         and for the main background processes of Drell-Yan, $t\bar{t}$, and diboson.}  
   	       \vspace*{1cm}
	       \label{fig:deta}
 \end{figure}

The dark photon signal events are simulated with the {\tt MadGraph5-aMC@NLO} Monte Carlo generator 
\cite{mg1,mg2,mg3}, and an already available Universal FeynRules Output (UFO) model 
\cite{ufoo} \footnote{\url{http://insti.physics.sunysb.edu/~curtin/hahm\_mg.html}}.
 The parton level events are passed through {\tt Pythia 8} \cite{pythi} to perform parton shower, hadronization and decay of
unstable particles. As mentioned, the effects of detector  are simulated with {\tt Delphes 3.3.2} package \cite{delphi}.
The simulated samples for background processes are also generated in a similar fashion.
The selection of events is designed to identify opposite-sign charged lepton events compatible with the dilepton events  
arising from dark photon, while suppressing the contribution of background processes. 
The analysis concentrates on events with prompt $Z_{D}$ decays and not on displaced dark photon.
In order to trigger the events, one can either rely on the single lepton or double lepton triggers.
The charged lepton transverse momentum requirements are applied to satisfy the double lepton triggers.
The leading lepton is required to 
have $p_{\rm T} > 25$ GeV and the second lepton must have $p_{\rm T} > 20$ GeV. Pseudorapidity
ranges are chosen to cover regions of good reconstruction quality.  For electrons and muons, it
is required that  $|\eta| < 2.4$. 
The leptons are required to satisfy  isolation criteria. The relative isolation for a charged lepton defined as:
\begin{eqnarray}
\label{relisolation}
\text{RelIso} = \frac{\sum_{k}^{\Delta R(\ell,k) < 0.3 }p_{\rm T}(k)}{p_{\rm T}(\ell)},
\end{eqnarray}
where in the numerator, the sum is over the transverse momenta of particles residing inside a cone
with a radius of $R = 0.3$ except the charged lepton. For both electrons and muons, it is required $\text{RelIso} < 0.15$.
In addition, $\Delta R(\ell^{+},\ell^{-})$ is required to be larger than $0.3$.
To reduce the contribution of low-mass resonances and charged leptons from hadrons decays, the invariant mass
of dilepton is required to be larger than 10 GeV. Furthermore, the transverse momentum of the lepton pair must
satisfy $p^{\ell\ell}_{\rm T} > 30$ GeV to reduce background contributions from nonprompt leptons.
In order to reduce background contributions from $VZ$ ($V = W,Z$) production,
events with a third, loosely-identified charged lepton with $p_{\rm T}  > 10$ GeV are discarded.
For further background suppression the magnitude of missing transverse momentum is required to be less than $20$ GeV
and events containing any b-jet with $p_{T} > 30$ GeV and $|\eta| <  2.4$ are discarded. This helps reduce the
contributions from $t\bar{t}$ and single top $tW$-channel.
The normalized distribution of $|\eta_{\ell^{+}}-\eta_{\ell^{-}}|$ for dark photon with $\epsilon = 0.01, m_{Z_{D}} = 200$ GeV
and for the main background processes after the above cuts are depicted in Figure \ref{fig:deta}.

In order to obtain the projected
sensitivity for the dark photon parameters, we define a $\chi^{2}$-statistic over the $|\Delta \eta|$ distribution as follows:
\begin{eqnarray}
\chi^{2}(\epsilon, m_{Z_{D}}) = \sum_{i \in bins} \frac{(N^{i}_{\rm SM+DP}(\epsilon, m_{Z_{D}}) - N^{i}_{\rm SM})^{2}}{(\delta^{i})^{2}},
\end{eqnarray}
where $N^{i}_{\rm SM+DP}(\epsilon, m_{Z_{D}})$ denotes the number of events in the $i$-th bin, after the selection
cuts described above, and $N^{i}_{\rm SM}$ is the SM prediction. The
uncertainties $\delta^{i}$ are both statistical and systematic uncertainties. 
Systematic uncertainties are taken from a CMS experiment analysis where differential cross section of Drell-Yan
has been measured in both electron and muon channels \cite{cmssys}. The systematic uncertainties are  considered 
differently in three mass regions of low mass ( below 40 GeV), $Z$-boson peak region, and high mass region (above 200 GeV).
In this work, on the electron and the muon channels, the systematic uncertainties related to the electron channel are applied. 
Since the uncertainties related to the electron channel are larger than the muon one, this makes the estimate of the sensitivities
more conservative.

\begin{figure}[t]
      	   \centering
   	   \includegraphics[width=0.7\linewidth]{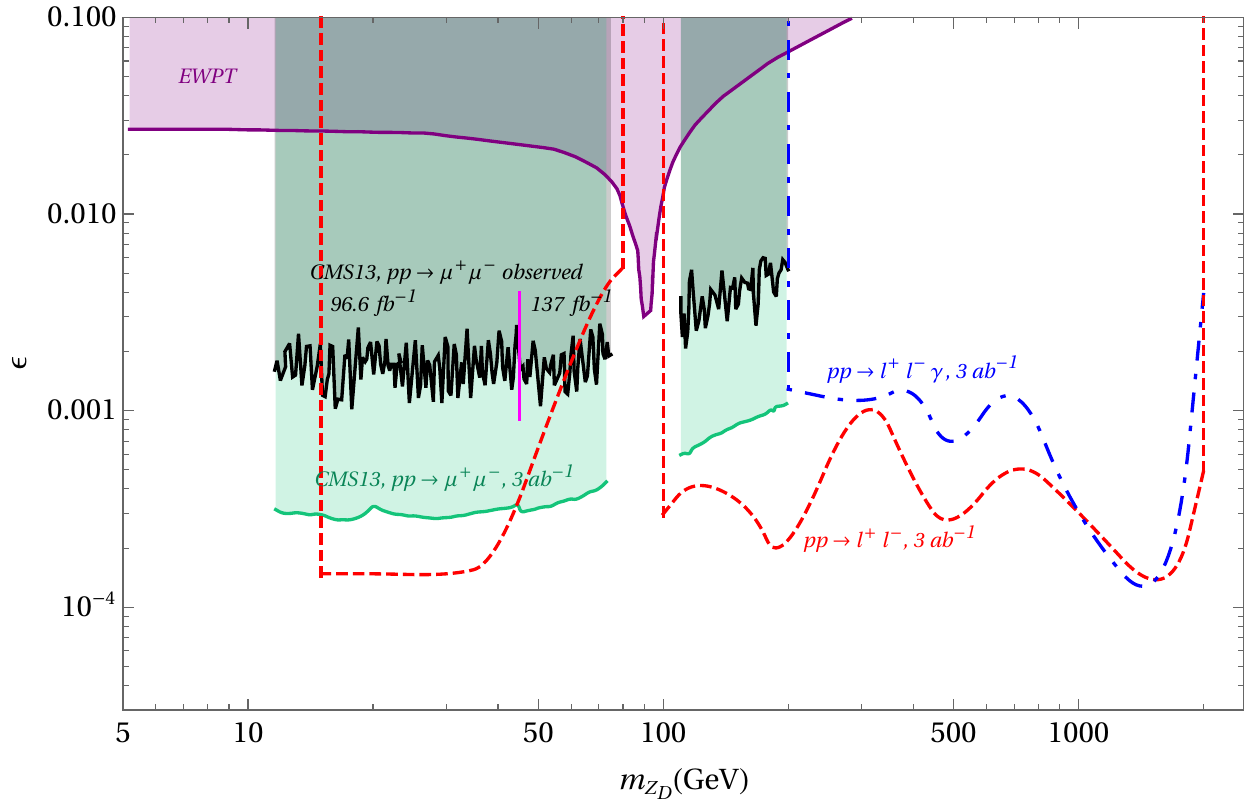}
	   \caption{ \small Future projection constraints on $\epsilon-m_{Z_{D}}$ parameters at $95\%$ CL
	   from dark photon Drell-Yan production and dilepton production in association with a photon
	   at the HL-LHC with 3 ab$^{-1}$ are depicted as red dashed and blue dot-dashed lines, respectively.
	   The black solid curve shows the observed upper limits on the kinetic mixing $\epsilon$ at $90\%$ CL as a function of 
	    the dark photon mass using dimuon final state in proton-proton collision data collected
	    by the CMS experiment at $\sqrt{s} = 13$ TeV with an integrated luminosity of 137 fb$^{-1}$ \cite{cmsmu}.  
	    The projection of the CMS experiment results from dimuon channel to an integrated luminosity of 3 ab$^{-1}$
	    is also presented as the green solid curve. Bounds at $95\%$ CL from the
          the electroweak observables are presented as purple shaded region\cite{curtin}. }
	   \label{DY_fig_epsilon}
   \end{figure}

Confidence level intervals are then calculated using:
\begin{eqnarray}
1-\text{CL} = \int_{\chi^{2}}^{\infty}f_{m}(x) dx,~ \chi^{2} = \chi^{2}(\epsilon, m_{Z_{D}}),
\end{eqnarray}
where $f_{m}(x)$ is the $\chi^{2}$ distribution of $m$ degrees of freedom evaluated considering the number of
free parameters and the number of total bins. In order to increase the sensitivity, an additional cut is 
imposed on the dilepton invariant mass $m_{\ell\ell}$ so that the best sensitivity is achieved. 
For each $m_{Z_{D}}$, it is required the lepton pair in each event to satisfy: $ \Delta_{1}  < |m_{\ell\ell} - m_{Z_{D}}| < \Delta_{2}$
where $\Delta_{1,2}$ are optimized  separately for each value of $m_{Z_{D}}$ to obtain the best limit on $\epsilon$.
The scan is performed using the $\chi^{2}$-statistic calculated at an
integrated luminosity of $\mathcal{L} = 3000$ fb$^{-1}$, and allowed region evaluated in the 
$\epsilon-m_{Z_{D}}$ plane at $95\%$ confidence level is depicted in Fig. \ref{DY_fig_epsilon}. 
For the $Z_{D}$ mass range of $15$ GeV to $2$ TeV (except $60-100$ GeV), any value of $\epsilon$ above 
$(1.4-10)\times 10^{-4}$ can be excluded using this analysis.

\subsection{Dark photon production associated with a photon at the LHC}
\label{sec:dygamma}
   
Within the SM, studying $Z$ boson production in association with a photon at the LHC plays an 
important role as it provides crucial tests of the electroweak sector. The $Z+\gamma$ measurements
are used to search for the effects of new physics beyond the SM such as direct couplings of $Z$ bosons to photons. 
There are measurements for the inclusive and differential $Z+\gamma$ production cross
sections using different datasets collected at center-of-mass energies and in $\ell^{+}\ell^{-}$, $\nu\nu$, and $b\bar{b}$ channels
by the ATLAS and CMS collaborations \cite{a1,a2,c1,c2}. 
Compared to the $b\bar{b}+\gamma$ and $\nu\nu+\gamma$ channels, the $\ell^{+}\ell^{-}+\gamma$ 
channel provides the possibility for cross section
measurements with lower background contributions and less uncertainties from systematic sources.
Therefore, in addition to dilepton production from dark photon, $\ell^{+}\ell^{-}\gamma$ channel 
with its clean final state and  under-control systematic sources can be used to probe dark photon physics.
In this section, the focus is on search for dark photon in $pp\rightarrow  \ell^{+}\ell^{-}\gamma$
considering a realistic CMS-like detector effects and the main sources of background processes.

The dominant background to the $ \ell^{+}\ell^{-}\gamma$ signal, 
comes from the SM $\ell^{+}\ell^{-}\gamma$ events. Other background contributions arise from
$t\bar{t}\gamma$ (with one or both top quarks decaying semileptonically), 
$VV$, $VV\gamma$ (where $V=W,Z$) (including $W$ and $Z$ bosons decays to 
final states involving $\tau$-leptons). 
There are contributions to the background composition from events containing Higgs boson with
$H\rightarrow Z+\gamma$ decays where $Z\rightarrow \ell^{+}\ell^{-}$ which is found to be negligible
after the full selection described below.

Similar to the previous analysis described in section \ref{sec:dy}, the signal and background 
processes are generated using  {\tt MadGraph5-aMC@NLO} event generator, then 
{\tt Pythia 8} to perform parton shower, hadronization and decay of
unstable particles and  the impact of a CMS-like detector is simulated with {\tt Delphes 3.3.2}.

Signal events are selected by requiring the existence of a photon together with a same-flavor lepton ($e$ or $\mu$) pair.
The contributions from background events 
coming from processes producing non-prompt leptons or photons are significantly suppressed
by applying isolation requirements on the photon candidate
and the two leptons. Similar trigger requirements and isolation criteria as described in 
the Drell-Yan analysis are imposed on the leptons. 
Photon candidate is required to have a pseudorapidity in the range $|\eta| < 2.5$ and to
have a transverse momentum greater than 20 GeV and the relative isolation $\text{RelIso} < 0.15$ as defined
in Eq.\ref{relisolation}. Events containing additional photons with $p_{T} > 20$ GeV, $|\eta| < 2.5$ and $\rm RelIso < 0.2$ are rejected.
In addition to the above requirements, photon candidates are required to
be well separated from all electrons and muons in the event by $\Delta R(\ell^{\pm},\gamma) > 0.3$, and
lepton candidates are required to be separated from each other in the event by
$\Delta R(\ell^{-},\ell^{+}) > 0.3$. The magnitude of missing transverse momentum vector
is required to be less than $20$ GeV which helps reduce the background containing 
neutrino(s) in the final state. To further suppress the contribution from $t\bar{t}+\gamma$, $VV\gamma$, and single top plus a photon,
events which have any jet with $p_{T} > 30$ GeV and
$|\eta| < 2.4$ are rejected.
For the sake of reducing the contribution of backgrounds from low-mass resonances, 
charged leptons from hadrons decays, misidentified photon, and $Z$ boson decays,  a lower cut is applied on the invariant mass
of dilepton. The concentration in this analysis is on the dark photon mass region above
$200$ GeV where the misidentified background contribution is suppressed. 
The signal efficiency for $m_{Z_{D}} = 400$  GeV and $\epsilon = 0.01$ is $2.0\%$.
The efficiencies of SM $\ell^{+}\ell^{-}\gamma$, $t\bar{t}\gamma$, $WW\gamma$, $ZZ\gamma$, $WZ\gamma$
are $1.97\%$, $0.006\%$, $0.51\%$, and $1.03\%$, $1.58\%$, respectively.

Furthermore, there are background events from Drell-Yan+jets process containing non-prompt photons,
for example arising from $\pi^{0}$ or $\eta^{0}$ decays. 
The probability for a jet misidentified as a photon varies with the misidentified photon transverse
momentum and is of the order of $\sim 10^{-3} (10^{-5})$ for low (high) $p_{T}$ jets \cite{atlas-fakerate}. Its contribution 
is estimated to be less than $3\%$ of the total background contribution in the signal region. We neglect this background
in the analysis nevertheless a dedicated and a detailed detector simulation has to be performed to estimate it.

In order to assess the sensitivity, having obtained the predicted signal,
a profile likelihood ratio test statistic is constructed over an 
optimised range of $m_{\ell\ell\gamma}$. The constraints on the kinetic mixing versus the dark photon mass at $95\%$ CL
are presented in Figure \ref{DY_fig_epsilon} as dot-dashed blue curve. 
An overall systematic uncertainty from Ref.\cite{zg}
is considered in extracting the limits.
The limit is obtained using 
an integrated luminosity of 3 ab$^{-1}$. As can be seen, in high mass region $m_{Z_{D}} \geq 200$ GeV,
kinetic mixing parameter $\epsilon$ can be excluded for any value above $\sim 10^{-4}$.

\subsection{Projection of  $Z_{D} \rightarrow \mu^{+}\mu^{-}$ measurement at HL-LHC}

In Ref.\cite{cmsmu}, CMS collaboration has performed a  search for a dark photon decaying to a pair of muons
$Z_{D} \rightarrow \mu^{+}+\mu^{-}$ using data recorded at a center-of-mass energy of 13 TeV. 
The search has been done in the mass ranges of $45-75$ GeV and $110-200$ GeV using  an
integrated luminosity of 137 fb$^{-1}$. It should be indicated that the search in the mass range of $11.5-45.0$ GeV 
has been carried out with 96.6 fb$^{-1}$ data collected using high rate dimuon triggers.
The result of this search is shown in Figure \ref{DY_fig_epsilon} as black solid curve. The limits obtained
using high rate dimuon triggers for the mass range of $11.5-45.0$ GeV is separated from 
those derived with the standard triggers by a vertical pink line. 
For the $Z_{D}$ mass range of $11.5-75$ GeV ($110-200$ GeV), any value of kinetic mixing parameter above 
$(1-2)\times 10^{-3} ((2-5)\times 10^{-3})$ has been excluded by this analysis. 
An extrapolation of the expected results to an integrated luminosity of 3 ab$^{-1}$ is performed
which are presented in Figure \ref{DY_fig_epsilon} as green solid curve. As seen, 
increasing the integrated luminosity to HL-LHC benchmark would improve the constraints
on kinetic mixing parameter $\epsilon$ by a factor of $\sim 6-7$.

We note that stronger limits from the Drell-Yan $Z_{D}$ production
using a shape analysis on $\Delta\eta$ distribution (red dashed curve in Fig.\ref{DY_fig_epsilon}) 
is achieved as compared to the extrapolation of CMS experiment results (green solid curve in Fig.\ref{DY_fig_epsilon}).
Different behaviour of $\Delta\eta$ distribution with respect to the SM background in particular for heavy dark photons is
the main reasons for this potential improvement. It is notable that performing the shape analysis on $\Delta\eta$ distribution
with only dimuon channel loosens the constraints around $15\%-25\%$ depending on the dark photon mass.
It is found that at large $Z_{D}$ masses $m_{Z_{D}} \gtrsim 800 $ GeV, the resulting limits from dark photon Drell-Yan
are comparable with those derived from $\ell^{+}\ell^{-}\gamma$ signature. 
In the dark photon mass region $200$ to $800$ GeV, $\ell^{+}\ell^{-}\gamma$ limits are slightly weaker than those of
$Z_{D}$ production.

\subsection{Prospects for a multi-TeV muon collider}
\label{sec:muon}

Muon colliders are in particular noticed due to several potential 
benefits such as availability of the full energy of muons in collisions,
clean environment compared to hadron colliders, and significant mass suppressed synchrotron radiation 
with respect to the electron-positron colliders. 
They are excellent options which allow us 
to scan the Higgs boson resonance and to accurately measure its mass and width. 
In addition, muon colliders are 
ideal to probe new physics effects beyond the SM and to study narrow resonances both as 
machines for precision and for exploratory purposes.  
The aim of this section is to present the potential of a future muon collider
in search for dark photon and  its ability to scan the dark photon parameter space.

We perform the search through the singly produced $Z_{D}$ in the s-channel via
Drell-Yan production, i.e. $\mu^{+}+\mu^{-} \rightarrow \ell^{+}+\ell^{-}$ with $\ell = e,\mu$. 
The sensitivity of dilepton measurements to dark photon can be obtained 
from dilepton mass spectrum measurement or other differential distributions.
Both in electron and muon colliders,  photon radiation is the
effect which needs to be considered when a narrow resonance is created 
in the annihilation channel. Such an effect has been important in observation
of $J/\psi$  in electron-positron collisions and 
in $Z$-boson production \cite{r1,r2,r3}.  Photon radiation in dark photon production at $\mu^{+}\mu^{-} $ collisions
leads to the following modification factor in the lowest order cross section: 
\begin{eqnarray}\label{reduction}
\left(\frac{\Gamma_{Z_{D}}}{m_{Z_{D}}}\right)^{\frac{4\alpha}{\pi}\log(\sqrt{s}/m_{\mu})},
\end{eqnarray}
where $\Gamma$ is the width of dark photon, $s$ is the center-of-mass energy 
of the collision, and $m_{\mu}$ is the muon mass.
Such  QED effects have been accurately calculated for the experiments at Large Electron Positron 
(LEP) collider up to two-loop corrections \cite{Nicrosini:1986sm}. As for the studies in muon colliders, these corrections
have been mentioned in studies related to Higgs boson line-shape and design of machine. For example, 
the impact of this modification factor on Higgs boson production rate is expected to be of the order of $50\%$ \cite{mumu}.
As seen in Eq.\ref{reduction}, this reduction factor depends on the center-of-mass energy 
so that the higher $\sqrt{s}$ the larger suppression rate.
For large dark photon masses compared to $Z$ boson $m_{Z_{D}} > m_{Z}$ and small values of
the strengths of kinetic mixing $\epsilon \ll 1$, $\Gamma_{Z_{D}}/m_{Z_{D}} < \Gamma_{Z}/m_{Z}$, where $\Gamma_{Z}/m_{Z}= 0.02$.
Therefore, in this region of parameter space ($m_{Z_{D}} > m_{Z}$ and $\epsilon \ll 1$),  
the suppression factor for Drell-Yan dark photon is greater than SM Drell-Yan and as a result
more  photon radiations are expected for SM Drell-Yan. This effect shows up 
for instance in  $|\vec{p}^{\ell\ell}_{T}|$ distribution as more photon radiations give rise to 
further imbalance in momentum of dilepton in the final state. In this section
based on the difference between the shape of SM Drell-Yan 
which is the main irreducible background and the dark photon signal in  $|\vec{p}^{\ell\ell}_{T}|$ distribution,
the sensitivity to parameter space is derived.

According to the dilepton final state, there are different SM background processes.
The SM $\ell^{+}\ell^{-}$ production,  $W^{+}W^{-}$ when both of the $W$ bosons decay 
leptonically, $ZZ$ when at least one of the $Z$ bosons decays to a lepton pair,  top quark pair in dilepton decay mode,
and $HZ$ when $Z\rightarrow \ell^{+}+\ell^{-}$ are the main background processes. 
The background processes are generated using {\tt MadGraph5_aMC@NLO} to produce hard events.
Showering, hadronization and decays of unstable particles are done with {\tt Pythia 8}.
The impact of detector is simulated using {\tt Delphes 3.5.0} considering the delphes card
for muon collider \footnote{\url{https://github.com/delphes/delphes/blob/master/cards/delphes_card_MuonColliderDet.tcl}}.  
The analysis is performed for two center-of-mass energies 1.5 and 6 TeV.

\begin{figure}[ht]
     	   \centering
	      	\includegraphics[width=0.5\linewidth]{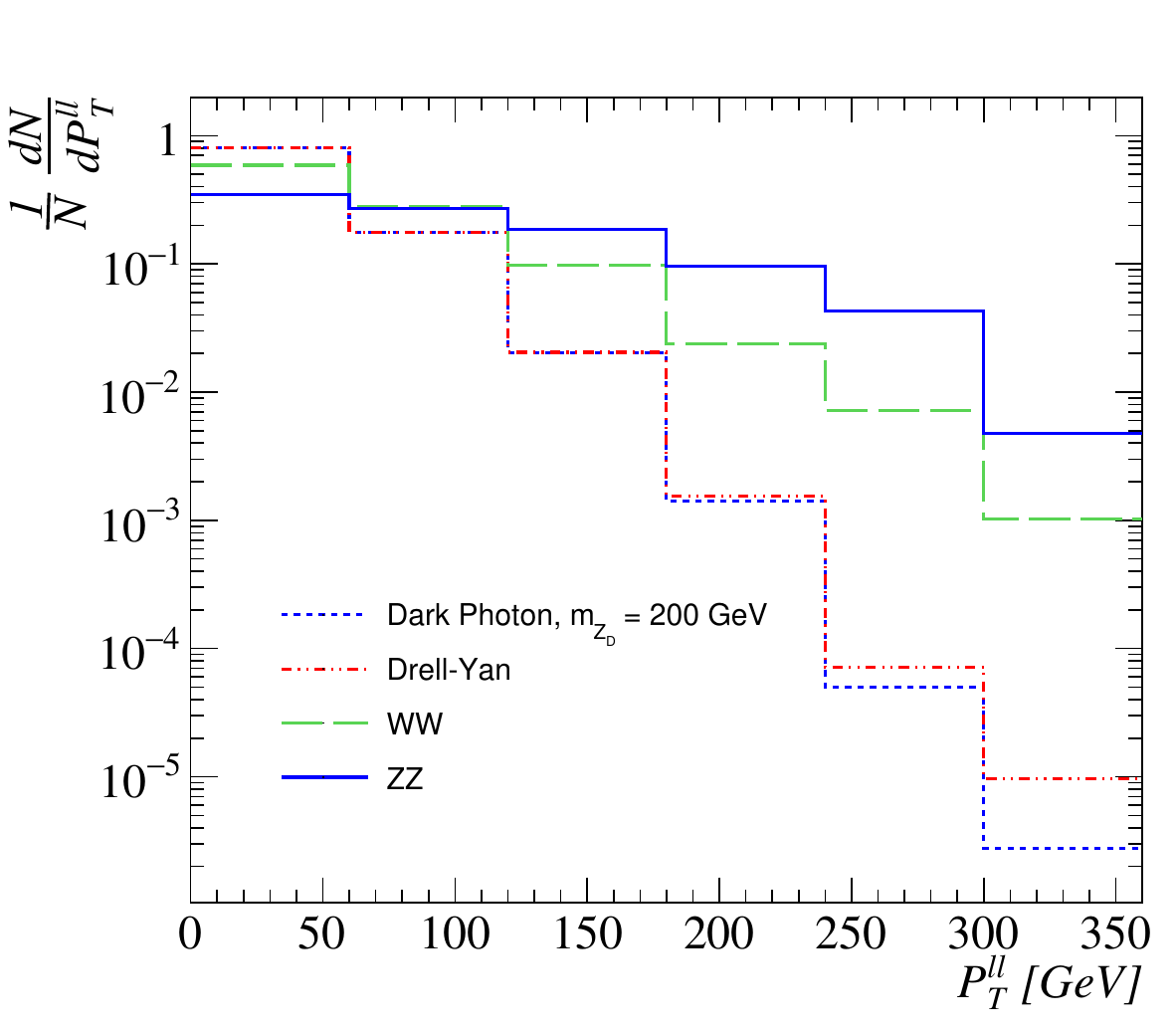}
   	       \caption{ \small Distributions of $|\vec{p}^{\ell\ell}_{T}|$ for dark photon signal with $m_{Z_{D}} = 200$ GeV
	       and the strength of kinetic mixing of $0.01$ at a center-of-mass energy of 1.5 TeV. The distributions for 
	       the main irreducible SM Drell-Yan background as well as $WW$ and $ZZ$ are depicted. }  
   	       \vspace*{1cm}
	       \label{fig:ptll}
  \end{figure}

The signal final state is composed of two oppositely charged same-flavour isolated leptons,
electrons or muons, compatible with a $Z_{D}$ boson decay. 
The  cross section of signal is much lower than that
of the major reducible and irreducible background processes, and therefore an optimum
 selection is needed to obtain a sample of sufficient purity. 
 To be consistent with the expected dark photon signal topology, the selection
requires leptons ($e,\mu$) with $p_{T} > 15$ GeV and  
 pseudorapidities of leptons are required to be $|\eta| < 2.5$ for electrons and $|\eta| < 3.0$
 for muons. Both electrons and muons have to be isolated by applying similar requirements on 
 the RelIso variable to the LHC analysis as defined in Eq.\ref{relisolation} presented in Sec. \ref{sec:dy}
 with a difference of the size of isolation cone which is taken to be $0.1$.
 In addition, to ensure the charged leptons are well isolated, it is required that $\Delta R (\ell^{+},\ell^{-}) > 0.3$.
 To reduce background processes containing neutrinos which appear as missing energy 
 in the final state the magnitude of missing transverse momentum is required to be less than
 20 GeV and the invariant mass of the dilepton system is required to be greater
 than $0.8\times \sqrt{s}$. These suppress $ZZ$, top quark pair, and $WW$ background remarkably.
As indicated, the signal topology is characterised by a dilepton system 
however due to photon radiations from initial state and/or final state, the momentum of the dilepton
system is not balanced and a deviation in total momentum is expected. 
In Figure \ref{fig:ptll},  the magnitude of the total momentum of the dilepton system in the transverse plane
is presented for signal with $m_{Z_{D}} = 200$ GeV
and the strength of kinetic mixing of $ \epsilon = 0.01$ and for 
the main irreducible SM Drell-Yan background, $WW$, and $ZZ$ processes. 
At $\sqrt{s} = 1.5$ TeV, the efficiency of signal for $m_{Z_{D}} = 200$  GeV and $\epsilon = 0.01$ is found to be $72.4\%$ and
the efficiencies for background processes SM Drell-Yan, $WW$, $ZZ$, top quark pair, $HZ$
are respectively $72\%$, $0.5\%$, $0.02\%$, $\lesssim 10^{-4}\%$, and $0.0$.

\begin{figure}[ht]
      	   \centering
   	   \includegraphics[width=0.6\linewidth]{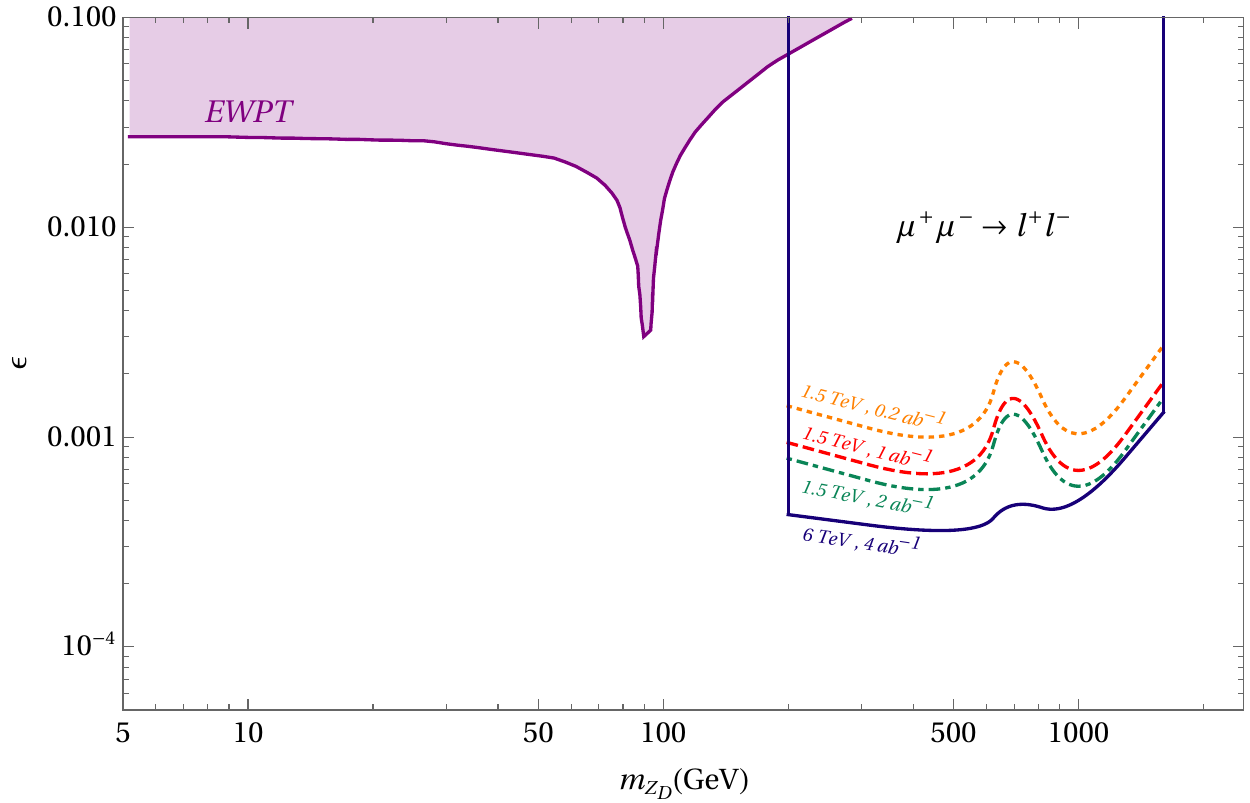}
	   \caption{\small Constraints on $\epsilon-m_{Z_{D}}$ plane at $95\%$ CL
	   from a muon collider at $\sqrt{s} = 1.5$ and $6$ TeV with the integrated luminosities
	   of (0.2,1,2) ab$^{-1}$ and 4 ab$^{-1}$, respectively.
	   Limits from the electroweak observables are given as purple shaded region at $95\%$ CL \cite{curtin}.}
	   \label{DY_muon}
   \end{figure}

As expected the distribution of $|\vec{p}^{\ell\ell}_{T}|$ for dark photon signal drops faster than 
the SM Drell-Yan. The distributions of $WW$ and $ZZ$ background processes have
a remarkable tail at large values of $|\vec{p}^{\ell\ell}_{T}|$ because of the presence of neutrinos
which cause larger imbalance in the total momentum of the dilepton system. 
For the signal process, the differential distribution for $|\vec{p}^{\ell\ell}_{T}|$ can be 
written in the following form:
\begin{eqnarray}
\frac{d\sigma(\epsilon,m_{Z_{D}})}{d|\vec{p}^{\ell\ell}_{T}|} = \frac{d\sigma_{\rm SM}}{d|\vec{p}^{\ell\ell}_{T}|}+
\epsilon^{2} \frac{d\sigma_{\rm int.}(m_{Z_{D}})}{d|\vec{p}^{\ell\ell}_{T}|}+\epsilon^{4}\frac{d\sigma_{Z_{D}}(m_{Z_{D}})}{d|\vec{p}^{\ell\ell}_{T}|},
\end{eqnarray}
where $d\sigma_{\rm int.}(m_{Z_{D}})/d|\vec{p}^{\ell\ell}_{T}|$ and $d\sigma_{Z_{D}}(m_{Z_{D}})/d|\vec{p}^{\ell\ell}_{T}|$
are the interference term and the pure dark photon contribution, respectively.
This allows us to scan the parameter space ($\epsilon, m_{Z_{D}}$) using a $\chi^{2}$-statistic on the $|\vec{p}^{\ell\ell}_{T}|$ distribution. 
The scan with the $\chi^{2}$-statistic calculated at the center-of-mass energies of $1.5$ TeV and
$6$ TeV is performed using the integrated luminosities of $0.2,1,2$ ab$^{-1}$ and $4$ ab$^{-1}$, respectively,  and contours evaluated in the 
($\epsilon, m_{Z_{D}}$) plane at $95\%$ confidence level  are shown in Figure \ref{DY_muon}.
The average luminosities for a future muon collider as suggested by 
the Muon Accelerator Program \cite{muc2} at $\sqrt{s} = 1.5$ TeV and $\sqrt{s} = 6$ TeV
are $1.25\times 10^{34}$ cm$^{-2}$s$^{-1}$ and $12\times 10^{34}$ cm$^{-2}$s$^{-1}$, respectively.
The integrated luminosity of $2$ ab$^{-1}$ at $\sqrt{s} = 1.5$ TeV is corresponding to 
five years of data taking and the benchmark integrated luminosity of $4$ ab$^{-1}$ at $\sqrt{s} = 6$ TeV
is taken from Ref.\cite{muc3}.

As seen in Figure \ref{DY_muon},  multi-TeV colliders at the center-of-mass energies of
1.5 TeV and 6 TeV offer sensitivity to $\epsilon \gtrsim 10^{-4}$ for dark photon with mass greater than
200 GeV to around 1 TeV. This is the same order of magnitude as exclusion limits achievable with
searches based on direct $Z_{D}$ production at the HL-LHC as presented in Figure \ref{DY_fig_epsilon}.
As a final comment, we discuss some possible ideas to improve the search strategy and boost the sensitivity. 
Extending the final state to the hadronic decays of the dark photon in addition to the dilepton final state
could be used as a good way to increase the signal statistics and improve the sensitivity. 
Exploiting multivariate techniques such as Boosted Decision Trees (BDTs) or Neural Networks (NNs)
by using effective discriminating variables help distinguish the dark photon signal from
the dominant background sources which would lead to obtain better results. 
Polarization of the muon beams can be used as a superior tool 
to probe dark photon and to handle background processes from signal
as they have different chiral structures.

  \section{Conclusions}
  \label{sec:conclusions}
  
Dark sector states show up in many extensions of the Standard Model which mainly
serve as candidates for the dark matter in the universe.
A dark sector can be proposed with an additional $\rm U(1)_{\rm D}$ dark gauge
symmetry. It can interact with the SM through kinetic mixing with the hypercharge gauge boson
where a kinetic mixing parameter $\epsilon$ deals with the coupling strength of the dark
photon and SM particles. 
In this paper, firstly we obtained bounds on dark photon parameters from partial wave
unitarity, examining the allowed $WW\rightarrow WW$ scattering processes in the limit of large center-of-mass energy. 
The contribution of dark Higgs in the $WW\rightarrow WW$ is considered. 
The limits from unitarity examination depends on the center-of-mass energy of the $WW$ scattering
and found to be $\epsilon < 0.001$ for $m_{Z_{D}} \leq 50$ GeV and $\sqrt{s} = 13$ TeV.
The results depend on the dark Higgs mass $m_{S}$ and its mixing angle $\theta_{h}$ with the SM Higgs boson
and are presented for small values of mixing angle $\theta_{h}$.
The bounds are not sensitive to $m_{S}$ and $\theta_{h}$ because the dark Higgs coupling with $WW$ 
is suppressed by $\sin\theta_{h}$.

The second part of the paper presented collider searches for dark photon.
The presence of dark photon could lead to deviations from the SM predicted total and differential
cross sections of processes like direct $Z_{D}$ and $Z_{D}$ production associated with a photon 
which are used in the present work.
These processes are golden channels for dark photon searches  providing that the
dark Higgs boson mixing with SM Higgs boson is small. 
Performing a fast detector simulation with {\tt Delphes} package and 
taking into account the main sources of background processes, scans were performed 
to constrain the dark photon parameter space using $pp\rightarrow Z_{D} \rightarrow \ell^{+}\ell^{-}$
and $pp \rightarrow \ell^{+}\ell^{-}+\gamma$. 
For $Z_{D}$ production, it has been shown that the presence of 
dark photon modifies the $|\Delta \eta| = |\eta_{\ell^{+}} - \eta_{\ell^{-}}|$ distribution
using analytical calculation.  The $|\Delta \eta|$ distribution
enables us to differentiate between signal and the main SM background processes.
With performing a $\chi^{2}$ fit on $|\Delta \eta| $ distribution, limits on $\epsilon$
versus $m_{Z_{D}}$ were derived at HL-LHC. Excluding $Z$ boson mass window, 
for dark photon mass from $15-80$ GeV and $200-2000$ GeV, $\epsilon$
could be excluded down to $(1.4-10)\times 10^{-4}$. 
In $\ell^{+}\ell^{-}+\gamma$ search, a test statistic is performed over the
$m_{\ell\ell\gamma}$ distribution to extract the limits. For $m_{Z_{D}} = 200$ GeV
to 2000 GeV, constraints between $0.0002-0.001$  obtained on $\epsilon$ at HL-LHC.
The $\ell^{+}\ell^{-}+\gamma$ channel is a complementary channel to $Z_{D}$ production at the LHC
in particular for large mass of dark photon where less background contribution contribute
in the spectrum.

Finally,  we showed that multi-TeV muon colliders with the clean environment
have excellent sensitivity to dark photons. 
Using $\mu^{+}+\mu^{-} \rightarrow \ell^{+}+\ell^{-}$ at $\sqrt{s} = 1.5$ and $6$ TeV, considering detector effects and
major sources of background, we showed that for $m_{Z_{D}}$ above $200$ GeV to 
around $1000$ GeV, $\epsilon$ can be probed down to almost $(3-5)\times 10^{-4}$.
Multi-TeV muon collider approaches the same sensitivity to dark photon as Drell-Yan process
at the HL-LHC. Our conclusions for the dark photon sensitivity were based on a preliminary detector
performance implemented in {\tt Delphes}, which may differ from the ultimate detector performance
of possible muon colliders.

\section*{Acknowledgement}
The authors are grateful to Gh. Haghighat for reading the manuscript and useful comments.

%
%

\end{document}